\documentclass{article}

\usepackage{arxiv}

\usepackage[utf8]{inputenc} 
\usepackage[T1]{fontenc}    
\usepackage{hyperref}       
\usepackage{url}            
\usepackage{booktabs}       
\usepackage{amsfonts}       
\usepackage{nicefrac}       
\usepackage{microtype}      
\usepackage{lipsum}		
\usepackage{graphicx}
\usepackage{natbib}
\usepackage{doi}
\usepackage{multirow}
\usepackage{multicol}
\usepackage{tabularx}
\usepackage{algorithm}
\usepackage{algorithmic}
\usepackage{lineno}
\usepackage{color}
\usepackage{listings}

\usepackage{newfloat}
\usepackage{listings}

\usepackage{booktabs}
\usepackage{multirow}
\usepackage[table,xcdraw]{xcolor}

\newcommand{\foo}{\textcolor{blue}{$\bullet$} \hspace{5pt}}

\lstdefinestyle{promptstyle}{
  backgroundcolor=\color{gray!10},
  aboveskip=3mm,
  belowskip=3mm,
  showstringspaces=false,
  columns=flexible,
  basicstyle={\tiny\ttfamily},
  numbers=none,
  stringstyle=\color{black},
  breaklines=true,
  breakatwhitespace=true,
  tabsize=3
}

\title{Investigating Vaccine Buyer's Remorse: Post-Vaccination Decision Regret in COVID-19 Social Media Using Politically Diverse Human Annotation}

\date{} 					

\author{
Miles Stanley\thanks{Equal contribution} \thanks{Work done at RIT} \\
University of Washington, Seattle \\
Washington \\
\texttt{milesds@cs.washington.edu} \\
\And
Soumyajit Datta\footnotemark[1]  \thanks{Soumyajit Datta and Ashiqur R. Khudabukhsh co-mentored Miles Stanley on the project.} \\
Rochester Institute of Technology \\
Rochester, NY 14623 \\
\texttt{sd3528@rit.edu} \\
\And
Ashutosh Kumar \\
Rochester Institute of Technology \\
Rochester, NY 14623 \\
\texttt{ak1825@rit.edu} \\
\And
Ashique KhudaBukhsh \footnotemark[3] \\
Rochester Institute of Technology \\
Rochester, NY 14623 \\
\texttt{axkvse@rit.edu} \\
}

\renewcommand{\shorttitle}{INVESTIGATING VACCINE BUYER’S REMORSE}

\begin{document}
\maketitle

\begin{abstract}
A significant gap exists in datasets regarding post-COVID-19 vaccination experiences, particularly “vaccine buyer's remorse”. Understanding the prevalence and nature of vaccine regret, whether based on personal or vicarious experiences, is vital for addressing vaccine hesitancy and refining public health communication. In this paper\footnote{This paper has been accepted at the AAAI 2026 AISI track.}, we curate a novel dataset from a large YouTube news corpus capturing COVID-19 vaccination experiences, and construct a benchmark subset focused on vaccine regret, annotated by a politically diverse panel to account for the subjective and often politicized nature of the topic. We utilize large language models (LLMs) to identify posts expressing vaccine regret, analyze the reasons behind this regret, and quantify its occurrence in both first and second-person accounts. This paper aims to (1) quantify the prevalence of vaccine regret; (2) identify common reasons for this sentiment; (3) analyze differences between first-person and vicarious experiences; and (4) assess potential biases introduced by different LLMs. We find that while vaccine buyer's remorse appears in only $<2\%$ of public discourse, it is disproportionately concentrated in vaccine-skeptic influencer communities and is predominantly expressed through first-person narratives citing adverse health events.
\end{abstract}


\section{Introduction}

The global COVID-19 vaccination campaign unfolded amidst a deeply polarized social landscape~\cite{khudabukhsh2021we} and a concurrent \textit{infodemic}, an overabundance of information, both accurate and misleading, that makes it difficult for people to find reliable guidance~\cite{nascimento2022infodemics}. This environment, fueled by the rapid circulation of confusing and often contradictory health content on social media, created fertile ground for post-decisional sentiments like vaccine regret to emerge and spread. While a substantial body of research has explored the drivers of pre-vaccination hesitancy, a significant gap remains in our understanding of post-vaccination sentiment, particularly the phenomenon of \textit{vaccine buyer's remorse}. Understanding the prevalence and nature of this vaccine regret is vital for refining public health communication, addressing the long-term erosion of institutional trust, and preparing for future health crises.

The concept of \textit{decision regret}, the distress an individual feels after making a health-related choice, is a well-established area of study \cite{becerra-perez2016more, zeelenberg1997consequences, brehaut2003validation}. Recent work has begun to explore this phenomenon in the context of COVID-19, linking the experience of adverse events to increased regret and a subsequent unwillingness to receive booster doses \cite{luo2022post}. Studies have also identified perceived coercion and disillusionment with vaccine efficacy as key drivers of this sentiment \cite{tayhan2025nursing}. This regret is often shaped by social media, where personal anecdotes about side effects can create negative expectations and amplify perceived negative experiences through the nocebo effect \cite{clemens2023social}, especially since compelling narratives can sway medical decisions even when presented alongside contradictory statistical data~\cite{line2024anecdotes}.

For decades, public health agencies have conducted post-market safety surveillance through formal channels like the Vaccine Adverse Event Reporting System (VAERS), a passive system that relies on voluntary reports from the public and clinicians \cite{shimabukuro2015safety}. In the digital age, social media platforms like YouTube have become vast, informal analogs to VAERS, hosting millions of unsolicited, user-generated accounts of personal and vicarious health experiences. Analyzing this discourse offers an opportunity to understand public sentiment at scale, yet it presents significant methodological challenges. Distinguishing first-person accounts from vicarious narratives, interpreting the nuanced emotion of regret, and mitigating potential model biases requires a sophisticated analytical approach.

To address these challenges, this study introduces a novel dataset and a multi-stage hybrid inference pipeline to analyze a large corpus of YouTube comments related to COVID-19 vaccination. 

\noindent\textbf{Contributions:} First, we focus on an underexplored  aspect of vaccine discourse on the social web -- vaccine regret or vaccine buyer's remorse. Second, we create a benchmark dataset for vaccine regret, annotated by a politically diverse panel of raters to account for the subjective and often politicized nature of the topic~\cite{dolman2023opposing}. Our dataset consists of 2,000 YouTube comments each annotated by three raters (overall, 201 unique raters)\footnote{Dataset, annotator details, and an expanded version with supplemental information (SI) are publicly available at \\\url{https://github.com/Social-Insights-Lab/Vaccine-Buyers-Remorse}.}. We develop and evaluate a computational pipeline to classify comments based on narrative perspective (first-person vs. vicarious) and the presence of regret. Finally, we use this pipeline to conduct a large-scale analysis of our corpus, guided by four primary research questions: (1) quantify the prevalence of vaccine regret; (2) identify the common reasons cited for this sentiment; (3) analyze the differences between first-person and vicarious expressions of regret; and (4) assess potential biases in our models' classifications.  

\begin{table*}[htb]
{
\scriptsize
\begin{center}

\begin{tabular}{|p{0.65\textwidth}|p{0.12\textwidth}|p{0.12\textwidth}|}
\hline
\textbf{Example Comment} & \textbf{Regret Label} & \textbf{Subject Label} \\
\hline

\cellcolor{gray!15} \textit{I got my first shot, then tried to get a mamogram and they will not do it until 4 weeks after my second shot.  apparently the limp nods have a chance of swelling.  I wouldn't have gotten the shot if i knew that.  they are not telling us everything.} 
& \cellcolor{gray!15} Positive  &  \cellcolor{gray!15}First-Person \\
\hline

\textit{i didn't vote for world shut down. i can do the research myself and have decided this particular vaccine is not safe for many reasons. my own doctor is no longer recommending it and regrets taking it himself.} 
& Positive & Third-Party \\
\hline

\cellcolor{gray!15} \textit{coming home after my booster shot, i just crossed paths with the coroner as he was removing a body from my apt. building - covid is everywhere.  get vaxed!} 
& \cellcolor{gray!15}Negative  & \cellcolor{gray!15}First-Person \\
\hline

\textit{isn't dying the activity you do after the vaccine} 
& Negative  & Unspecified \\
\hline
\end{tabular}
\end{center}

\caption{Examples of manually annotated comments from the dataset.}
\label{tab:example_comments}
}
\end{table*}

\section{Related Work}
Our work is situated at the intersection of public health, psychology, and natural language processing. We draw upon existing literature in three primary areas: the psychological underpinnings and real-world manifestations of vaccine regret, the application of large language models (LLMs) to analyze health-related social media data, and the broader context of public health challenges in the digital age.


Methodologically, our work leverages LLMs to move beyond traditional sentiment analysis, which is often insufficient for capturing the complex and often ambivalent emotion of regret. Traditional sentiment analysis often fails to capture the complexity of user opinions, such as sarcasm or mixed emotions \cite{liu2022dual}. Recent studies have demonstrated that LLMs can perform more sophisticated, multi-layered sentiment and topic analysis, identifying not only positive or negative sentiment but also discrete emotions and their underlying drivers \cite{yin2024unlocking}. Similar work has used deep learning models to monitor public opinion and extract reported side effects from Twitter \cite{portelli2022monitoring}, \cite{jain2025twitter}. LLMs have proven effective in specific information extraction tasks, such as identifying adverse events following vaccination from social media posts with high precision \cite{li2025enhancing}, \cite{sehgal2025conversations} and \cite{zhou2025ai}. However, the literature also cautions that LLMs are not without limitations; their performance is highly dependent on effective prompt engineering, and they can be prone to factual inaccuracies and inherent biases, necessitating careful validation against human-annotated data \cite{he2024using, crowl2025measuring}.

Prior work also examines broad vaccine debates, latent arguments, and anti-vaccine themes, but none focus on post-vaccination regret. \cite{pacheco2022holistic,pacheco2023interactive} analyze vaccine narratives and concept learning; \cite{islam2025uncovering,islam2025discovering, islam2022understanding} study campaign messaging and LLM-assisted theme discovery; and \cite{wawrzuta2021arguments} analyze anti-vaccine arguments. Our paper uniquely targets \textit{vaccine buyer’s remorse}, distinguishing first- vs third-person narratives and quantifying regret prevalence using a novel benchmark dataset.  

Finally, our annotation strategy is grounded in recent work addressing annotator subjectivity and bias. Labeling  politically charged content is inherently subjective. \cite{weerasooriya2023vicarious} demonstrated that annotators' political beliefs systematically influence how they perceive and label potentially offensive content. They found that recruiting a politically diverse panel of annotators is a crucial step in understanding and accounting for these perceptual differences. Our study employs a similar methodology as in \cite{crowl2025measuring,pofcher2025hope} to create a benchmark dataset that explicitly accounts for political diversity.

\section{Dataset}
\label{sec:dataset}

\subsubsection{Data Collection:} We curate a dataset $\mathcal{D}_\textit{newspool}$ of 80,307,930 comments posted on 65886 YouTube videos ($\mathcal{V}_\textit{newspool}$) from the official channels of three major U.S. cable news networks: Fox News, CNN, and MSNBC via the official \texttt{YouTube Data API v3}.  This
dataset has found prior use in studying  election-related discourse~\cite{khudabukhsh2021we,Capitol}, health-related discourse~\cite{yoo2023auditing}, LGBTQ+ discourse~\cite{pofcher2025hope}, rater subjectivity~\cite{weerasooriya2023vicarious,DBLP:conf/emnlp/PanditaWDLRKZH24,ARTICLEAAAI2025}, and political identity projections~\cite{PoliticlaIdentityProjections}. This selection was made to capture a spectrum of potential political viewpoints among commenters, a key factor in exploring variations in sentiment. The comments for the videos selected for inclusion were published within the timeframe from 14th December 2020 to 31st October 2024 and feature all types of content such as daily briefings, news updates, or official announcements but not particularly related to COVID-19. Along with these mainstream sources, we also included 981 videos ($\mathcal{V}_\textit{influ}$) with 847,702 comments ($\mathcal{D}_\textit{influ}$) from prominent YouTube influencers taking part in the vaccine discourse within the same timeframe. This addition was made with the understanding that many individuals now consume news from a variety of online creators~\cite{zimmermann2023political}, and the discussions in their comment sections warrant similar scholarly attention. 

Two independent reviewers with a consensus
manually categorized influencer channels based on a qualitative review of their content. We define channels as ``\textit{pro-vaccine}" if their content consistently aligned with public health guidance and encouraged vaccination. Conversely, channels were defined as ``\textit{vaccine-skeptic}" if they frequently questioned the safety or efficacy of vaccines, focused on adverse events, or expressed opposition to vaccine mandates.

\paragraph{Identifying Vaccine-Relevant Comments:}To build our corpus, we first performed an initial, keyword-based filtering of all collected comments to isolate those relevant to vaccination. This process identified comments containing keywords related to personal experiences, side effects, and potential regret. After the filtration, we got total comments of 1,370,101 ($\mathcal{D}_\textit{news})$ from 54,666 videos ({$\mathcal{V}_\textit{news}$}).

\paragraph{Constructing the Benchmark and In-the-wild Dataset:} From the filtered comments dataset $\mathcal{D}_\textit{news}$, we construct a dataset designed to categorize content based on sentiments and perspectives related to vaccine regret. As the \textit{Positive for Regret} class is considerably less common than irrelevant or neutral comments, a simple random sampling for our benchmark dataset would be inefficient. To address this, we employ a multi-faceted approach to identify comments more likely to be valuable for human annotation. This involved using a combination of simple regular expressions alongside zero-shot and few-shot prompting with an LLM to classify a large, unseen pool of comments. This process enabled us to purposefully sample comments identified as likely \textit{Positive for Regret}, creating a set for our full crowd-sourced annotation study and ensuring a more balanced and functional benchmark dataset $\mathcal{D}_\textit{bench}$ comprising 2,000 comments. For classification task we divide $\mathcal{D}_\textit{bench}$ into a training split of 80\% of our benchmark dataset ($\mathcal{D}_\textit{train}$), with the remaining portion reserved for testing ($\mathcal{D}_\textit{test}$). After validation, we processed a total of 600,000 comments. This corpus was balanced between mainstream news sources (300,000 comments, with 100,000 from each of Fox News, CNN, and MSNBC) and influencer channels (300,000 comments, with 150,000 from pro-vaccine influencers and 150,000 from vaccine-skeptic influencers). These comments were categorized as in-the-wild dataset ($\mathcal{D}_\textit{wild}$).

The annotation scheme captures two key dimensions:

\noindent\foo\textbf{\textit{Vaccine Regret:}} This dimension categorizes comments into one of two classes: \textit{Positive for Regret} and \textit{Negative for Regret}. The \textit{Positive for Regret} class includes both \textit{Explicit Regret} (direct statements like ``I regret taking the vaccine'') and \textit{Implicit Regret} (statements strongly suggesting dissatisfaction, e.g., ``I wish I never got it, my health has been terrible since''). The \textit{Negative for Regret} class includes comments that are unrelated to the topic or are neutral statements about an individual's vaccination status.

\noindent\foo\textbf{\textit{Narrative Perspective:}}{} For comments indicating regret, this dimension identifies the narrative point of view. Categories include \textit{First-Person} (personal experience), \textit{Third-Party/Vicarious} (reporting another specific individual's knowledge), and \textit{Unspecified}. We only categorized regret for comments that were from a first or third party, not unspecified (which includes general statements, e.g., ``People regret taking this vaccine").


\subsubsection{Annotation Study Design.}
We conduct a crowd-sourced annotation study designed to mitigate potential political bias in subjective annotations to generate our $\mathcal{D}_\textit{bench}$ labels.

\noindent\textbf{\textit{Annotator Recruitment:}} Following the methodology of \cite{weerasooriya2023vicarious}, \cite{crowl2025measuring}, and \cite{pofcher2025hope}, we recruited a panel of annotators with diverse, self-identified political affiliations: Republican, Democrat, and Independent.\\ 
 \noindent\textbf{\textit{Annotation Process:}} Each comment in the $\mathcal{D}_\textit{bench}$ was independently annotated by one annotator from each of the three political groups.\\
\noindent\textbf{\textit{Disagreement Resolution:}} Prior literature has considered diverse approaches to resolving inter-annotator disagreements (e.g., majority voting~\cite{davidson2017automated,wiegand2019detection} or third objective instance~\cite{DBLP:conf/ranlp/GaoH17}). The final label for each comment on each dimension was determined by a majority vote. 

\subsubsection{Annotation details:} We used Prolific\footnote{Prolific: https://www.prolific.com} to recruit annotators, while hosting the annotation questionnaire on our custom-built platform. A total of 2,000 comments were divided into 67 batches of 30, with 201 unique annotators participating in the labeling process. The annotator pool was evenly distributed across political affiliations (Democrat, Republican, Independent), with a near-equal gender ratio (106 male, 95 female) and an average age of 42.5 years. All annotators were US citizens. The median annotation time was $\sim$19 minutes. Each annotator was compensated \$4 per batch, estimated for a 30-minute task.



\begin{table*}[htb]
\centering
\scriptsize
\begin{tabular}{|l||cc|cc|cc||cc|cc|cc|}
\hline
\multirow{2}{*}{\textbf{Model}} &
\multicolumn{6}{c||}{\textbf{Zero-Shot}} &
\multicolumn{6}{c|}{\textbf{Few-Shot}} \\
\cline{2-13}
& \multicolumn{2}{c|}{Subject} & \multicolumn{2}{c|}{Vaccinated} & \multicolumn{2}{c||}{Regret}
& \multicolumn{2}{c|}{Subject} & \multicolumn{2}{c|}{Vaccinated} & \multicolumn{2}{c|}{Regret} \\
& F1 & Accuracy & F1 & Accuracy & F1 & Accuracy & F1 & Accuracy & F1 & Accuracy & F1 & Accuracy \\
\hline

\texttt{mistral-small3.2:24b} 
& 0.734 & 0.746 & 0.824 & 0.828 & 0.739 & 0.757 
& 0.695 & 0.702 & 0.836 & 0.837 & 0.790 & 0.816 \\

\texttt{gpt-4o-mini} 
& 0.672 & 0.674 & \textbf{0.842} & \textbf{0.843} & 0.803 & \textbf{0.855} 
& 0.647 & 0.650 & 0.839 & 0.839 & 0.804 & \textbf{0.853} \\

\texttt{mixtral:8x22b} 
& \textbf{0.745} & \textbf{0.769} & 0.770 & 0.783 & \textbf{0.806} & 0.842
& 0.733 & 0.743 & 0.833 & 0.836 & \textbf{0.808} & 0.846 \\

\texttt{llama3.1:70b} 
& 0.744 & \textbf{0.769} & 0.795 & 0.803 & 0.775 & 0.807 
& 0.709 & 0.717 & \textbf{0.840} & \textbf{0.841} & 0.800 & 0.843 \\

\texttt{qwen2.5:7b} 
& 0.617 & 0.620 & 0.793 & 0.794 & 0.766 & 0.796 
& 0.587 & 0.598 & 0.803 & 0.804 & \textbf{0.808} & 0.841 \\

\texttt{mistral:7b} 
& 0.683 & 0.702 & 0.763 & 0.771 & 0.767 & 0.795 
& 0.655 & 0.665 & 0.813 & 0.813 & 0.783 & 0.816 \\

\texttt{gemma3:12b} 
& 0.736 & 0.760 & 0.773 & 0.783 & 0.641 & 0.648 
& \textbf{0.743} & \textbf{0.763} & 0.796 & 0.803 & 0.687 & 0.698 \\

\texttt{llama3.1:8b} 
& 0.681 & 0.715 & 0.750 & 0.763 & 0.729 & 0.748 
& 0.649 & 0.667 & 0.792 & 0.794 & 0.774 & 0.807 \\

\hline
\end{tabular}
\caption{Performance comparison across models in zero-shot and few-shot settings on the held-out test set}
\label{tab:zero_few_shot}
\end{table*}

\section{Methodology and Experiment Design}
\label{sec:experimental_design}

\subsubsection{Zero and Few Shot Classification:}
We select eight diverse and widely-used LLMs (both open and closed source and ranging from 7B to 70B in size): \texttt{mistral-small3.2:24b}~\cite{mistral2024small32}, \texttt{mistral:7b}~\cite{Jiang2023Mistral7}, \texttt{mixtral:8x22b}~ \cite{Jiang2024MixtralOE}, \texttt{llama3.1:8b} ~\cite{grattafiori2024llama}, \texttt{gemma3:12b} ~\cite{team2025gemma}, \texttt{qwen2.5:7b} ~\cite{bai2023qwen}, \texttt{llama3.1:70b} ~\cite{grattafiori2024llama}, and \texttt{gpt-4o-mini} ~\cite{hurst2024gpt}. We perform zero-shot classification  on $\mathcal{D}_\textit{bench}$ using default hyperparameters with enforced JSON outputs, following best practices from \cite{ziems2023can}, to assess the models’ out-of-the-box reasoning ability without task-specific tuning. We also evaluate $\mathcal{D}_\textit{bench}$ on a few-shot setting \cite{brown2020language} to understand whether minimal supervision improves performance and consistency across models.
\subsubsection{Supervised Classification:}For supervised classification, we finetune three models with varying architectures \texttt{llama3.1:70b}, \texttt{llama3.1:8b}, and \texttt{mixtral:8x7b} using a LoRA-based approach \cite{hu2021lora} on $\mathcal{D}_\textit{train}$. This method allows us to assess whether the models can better understand and adapt to the classification task through task-specific supervision. We evaluate their performance on $\mathcal{D}_\textit{test}$ using standard metrics: precision, recall, F1-score, and accuracy on a held-out validation set.

\subsubsection{Multi-Stage Hybrid Inference Pipeline:}To classify a large volume of user comments with both nuance and efficiency, we designed a two-stage hybrid inference pipeline to balance computational cost and accuracy: 
\paragraph{Stage 1: Relevance Filter:}
The first stage acts as a high-throughput \textit{Relevance Filter}, using a Natural Language Inference (NLI) model from \cite{sileo2024tasksource}, \texttt{ModernBERT-large-nli}. This model determines if a comment is relevant to the topic of vaccines by treating the comment as a premise and evaluating its entailment with a specific hypothesis. We tuned this stage by testing 24 combinations of different hypotheses and acceptance thresholds on our validation set. The best-performing configuration, which achieved an F1-score of 0.8680, utilized the hypothesis: \textit{``This comment mentions or discusses anything related to vaccines, vaccination, or immunization"} with an acceptance threshold of 0.01. This initial filtering step efficiently removes a large volume of irrelevant comments, ensuring our more computationally intensive model is reserved for relevant data and reducing the likelihood of the LLM producing off-task or malformed responses.
\paragraph{Stage 2: Expert Reasoner:}
Comments that pass the relevance filter proceed to the second stage, which uses a fine-tuned model. This LLM performs a multi-label classification in a single pass, identifying the subject (self, other, or unspecified), vaccinated status, and regret status using a detailed prompt with specific rules and examples.

\section{Results and Discussions}

\subsection{Classification Task}

Table~\ref{tab:zero_few_shot} shows that larger models generally perform better across tasks, with \texttt{mixtral:8x22b} achieving the highest F1 score for Regret classification in both zero-shot and few-shot settings; \texttt{gpt-4o-mini} leads in Vaccinated prediction (zero-shot), while surprisingly \texttt{gemma3:12b} performs strongly on Subject in few-shot. Few-shot setups yield marginal improvements for most models, indicating that even minimal supervision can enhance performance on complex social classification tasks.

Table~\ref{tab:finetuned_comparison} shows the result for our finetuned models and it improves on zero and few shot metrics. While the \texttt{mixtral:8x7b} \cite{jiang2024mixtral} model showed the highest performance on the Subject classification task, the \texttt{llama3.1:70b} model demonstrated the strongest and most consistent performance on the two tasks most central to our RQs by achieving the highest F1-score and accuracy for identifying Furthermore, in our practical testing, we observed that the Llama model had a faster inference speed than the Mixtral model. Given its performance on key tasks and its computational efficiency, we selected \texttt{llama3.1:70b} as the most suitable model for our final inference pipeline.

\subsubsection{Pipeline Performance:}
Our two-stage hybrid inference pipeline was evaluated on the held-out test set of 400 comments from our benchmark dataset, The pipeline achieved an overall Exact Match (requiring all fields to be correct) of 62.00\%. The model demonstrated strong performance in identifying vaccinated (F1(macro)=0.87) and regret (F1(macro)=0.82) status, though it faced more challenges in the three-way subject classification (81.50\% accuracy \& F1(macro)=0.8), particularly with the unspecified class (F1(macro)=0.67).

\subsubsection{Pipeline Processing for $\mathcal{D}_\textit{wild}$:}
The first stage of our pipeline, the NLI relevance filter, identified 243,547 (40.6\%) of the comments as relevant to the topic of vaccines. The relevance rate was considerably higher for mainstream news sources (52.9\%) than for influencer channels (28.3\%).


\begin{table*}[htbp]
\centering
\scriptsize

\begin{tabular}{|l||cc|cc|cc|c|}
\hline
\multirow{2}{*}{\textbf{Model}} &
\multicolumn{6}{c|}{\textbf{Fine-Tuned Performance}} & \multirow{2}{*}{\textbf{Inference Speed}} \\
\cline{2-7}
& \multicolumn{2}{c|}{Subject} & \multicolumn{2}{c|}{Vaccinated} & \multicolumn{2}{c|}{Regret} & \\
& F1 (macro) & Accuracy & F1 (macro) & Accuracy & F1 (macro) & Accuracy & (sec/400 comments) \\
\hline

\texttt{llama3.1:70b} 
& 0.77 & 80.25\% & \textbf{0.87} & \textbf{87.00}\% & \textbf{0.83} & \textbf{87.00}\% & 650 \\

\texttt{llama3.1:8b} 
& 0.72 & 76.00\% & 0.86 & 85.75\% & 0.81 & 84.50\% & 164 \\

\texttt{mixtral:8x7b} 
& \textbf{0.79} & \textbf{81.50}\% & 0.86 & 85.75\% & 0.83 & 87.25\% & 2,703 \\

\hline

\end{tabular}
\caption{Performance and inference speed comparison across fine-tuned models on the held-out test set.}
\label{tab:finetuned_comparison}
\end{table*}

\subsection{In-the-wild results}

\subsubsection{Prevalence of Vaccine Regret Across Sources:}
\begin{figure}[htb]
    \centering
    \includegraphics[width=0.8\textwidth]{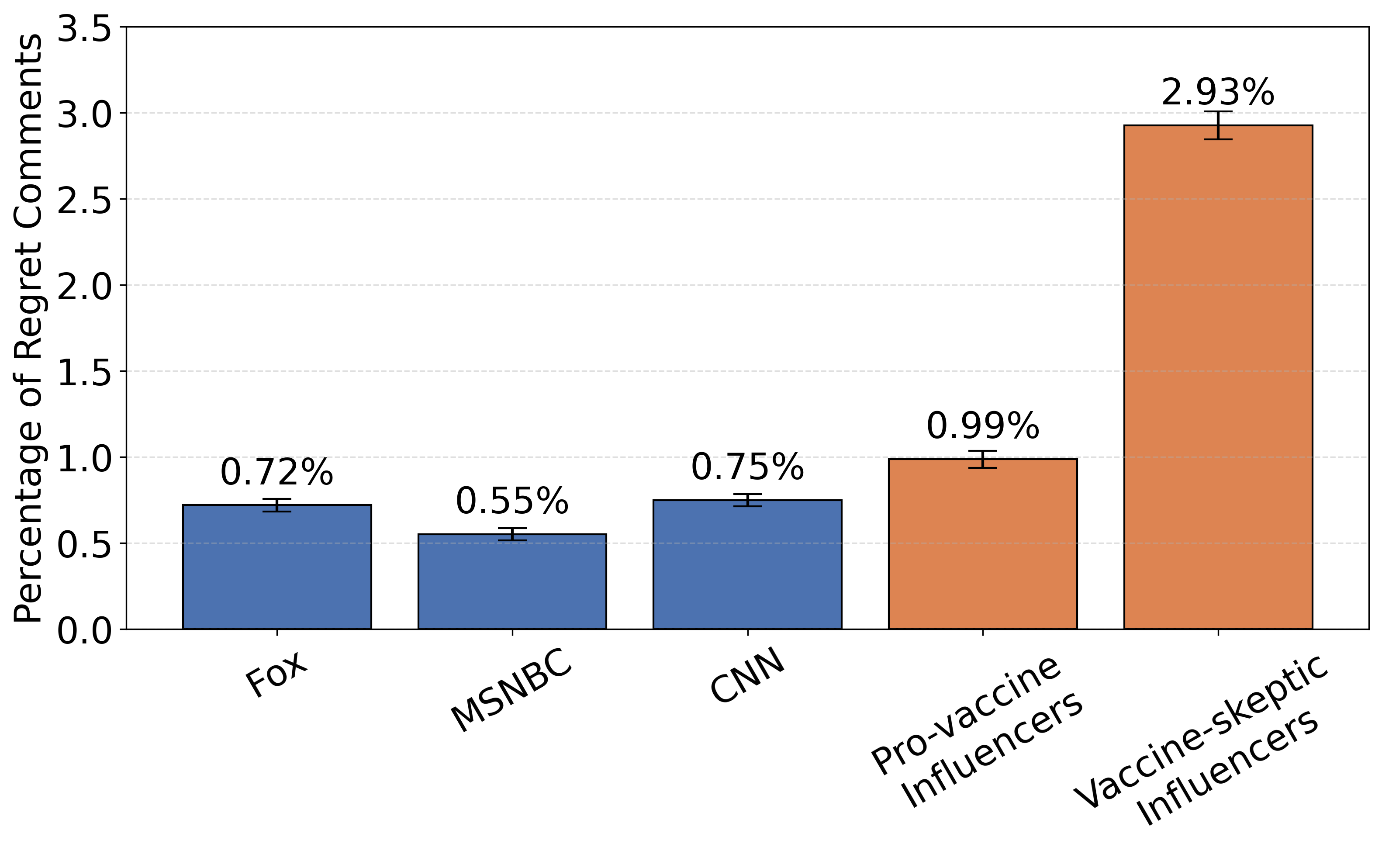}
    \caption{Percentage of regret comments across news sources and influencer categories.}
    \label{fig:regret_by_source}
\end{figure}

From the pool of 243,547 relevant comments, the pipeline identified 2,727 (1.1\%) as expressing regret. As shown in Figure~\ref{fig:regret_by_source}, the rate of regret varied significantly across source categories. The overall rate of regret was significantly higher on influencer channels (1.9\%) than on mainstream news channels (0.7\%) ($\chi^2$=780.29, p $<$ 0.001). Within the influencer category, the difference was even more stark: vaccine-skeptic channels exhibited a regret rate of 2.9\%, a statistically significant difference from the 1.0\% rate on pro-vaccine channels ($\chi^2$=421.10, p $<$ 0.001). In contrast, the variations in regret rates among the three mainstream news outlets were also statistically significant, though less pronounced ($\chi^2$=16.23, p $<$ 0.001).

Our analysis reveals that while expressions of vaccine regret are a persistent theme in online discourse, they appear in only 1.1\% of relevant comments. This provides a quantitative anchor to a topic often dominated by powerful anecdotes, which are known to impact medical decisions even when presented alongside statistical data~\cite{line2024anecdotes}. This suggests that while vaccine regret is a salient narrative, its actual prevalence in this discourse is far lower than its potential amplification within the broader \textit{infodemic} might suggest.

The prevalence of regret, however, is not uniform across online communities. The rate of regretful comments on influencer channels was more than double that on mainstream news channels, with vaccine-skeptic influencers hosting a rate nearly three times higher than their pro-vaccine counterparts. This aligns with prior work on the \textit{infodemic}~\cite{nascimento2022infodemics} and the significant role of online creators in shaping public opinion and health discourse~\cite{zimmermann2023political}. These channels may foster echo chambers where expressions of regret are more common, normalized, and amplified.

\subsubsection{Analysis of Narrative Perspectives:}
Of the 2,727 comments expressing regret, the majority (67.9\%) were first-person (self) narratives. This indicates that individuals sharing their own personal stories are the primary source of regretful sentiment in these online spaces.

\subsection{Substantive Findings}

\subsubsection{Analysis of Vicarious Relationships}
To further understand the social dynamics of vicarious regret narratives, we perform an additional classification on comments identified by the pipeline as having subject \textit{other}.

The core categories for this task are directly informed by empirical research on the social networks that influence vaccination decisions, which identifies a hierarchy of influential relationships \cite{brunson2013impact}. During our preliminary review of the data, we also observed that a number of comments referenced the experiences of celebrities, politicians, and other well-known individuals. This type of parasocial, one-to-many influence does not fit into the interpersonal categories, so we added a Public Figure category to capture this distinct form of vicarious narrative. The final categories are: Spouse or Partner, Family Member, Friend, Health Care Provider, Public Figure, Other Acquaintance, and Unspecified. 

This classification is performed using a zero-shot prompting approach with the \texttt{Llama-3.1:70B-Instruct} \cite{llama3-2024} model 
and is validated by manually verifying the model's output on a random sample of comments.
We ran this on the author's relationship to the subject in all 875 comments in $\mathcal{D}_\textit{wild}$ identified as \textit{“other"}. This approach was first validated on a manually annotated set, where it achieved 90.71\% accuracy. 

\begin{figure}[htb]
    \centering
    \includegraphics[width=0.8\textwidth]{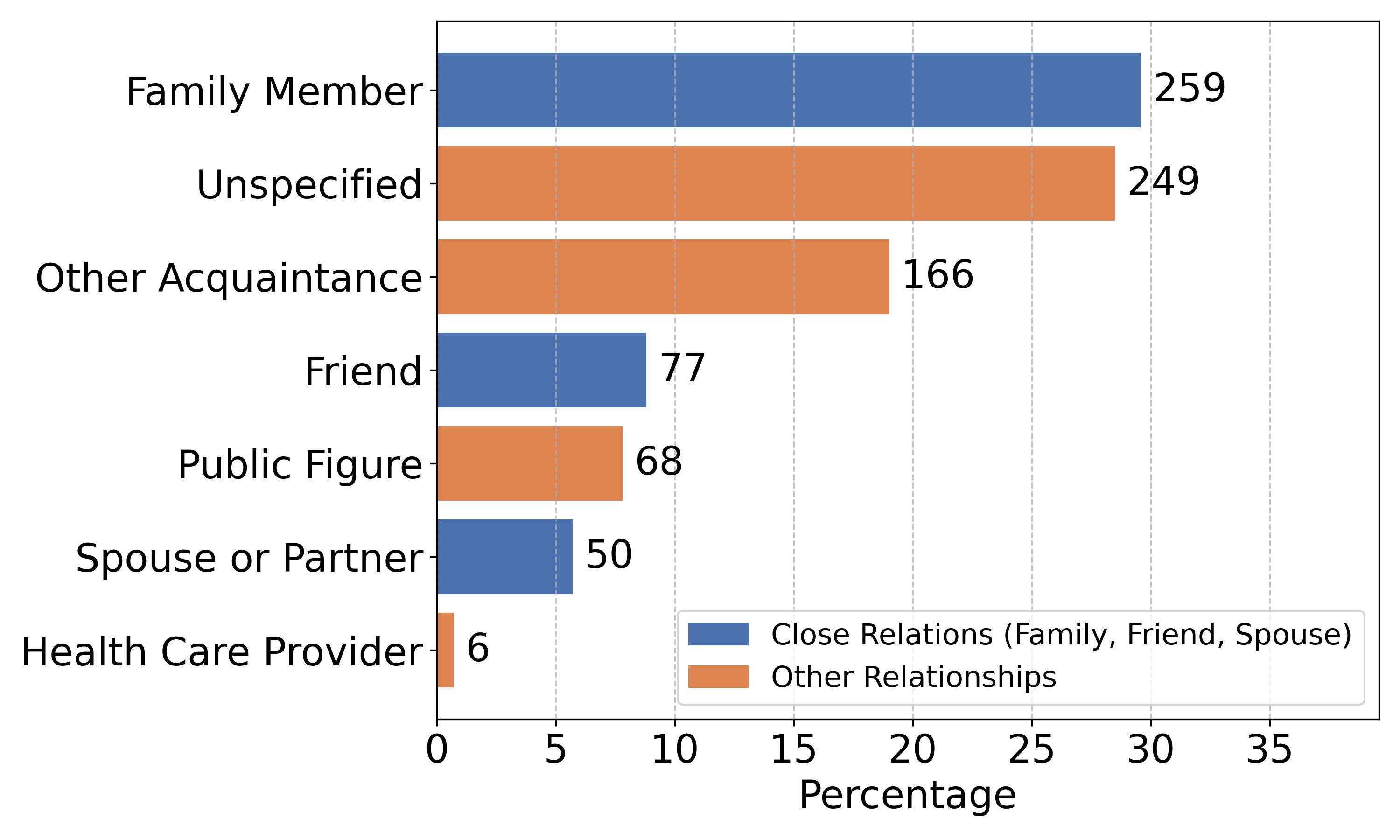}
    \caption{Overall Distribution of Relationships in Vicarious Regret Comments}
    \label{fig:relationship_overall}
\end{figure}

The analysis showed that \textit{Family Member} was the most frequently cited relationship (29.6\%), followed by \textit{Unspecified} (28.5\%) (see Figure~\ref{fig:relationship_overall} for full distribution). This underscores the role of intimate social networks in the dissemination of health narratives, as established by~\cite{brunson2013impact}.

\subsubsection{Analysis of Regret Reasons:}
\label{sec:analysis_reasons}
To identify the primary themes driving vaccine regret, we developed a set of categories informed by existing literature and our own data. The categories \textit{Adverse Health Event}, \textit{Perceived Coercion}, and \textit{Shift in Beliefs} are directly supported by prior qualitative research \cite{tayhan2025nursing, luo2022post}. In addition, we included the \textit{Lack of Efficacy} category after observing a prevalent theme of individuals expressing regret because they believed the vaccine did not work as promised (e.g., they still contracted COVID-19). We then employed a zero-shot prompting approach with the \texttt{Llama-3.1:70B-Instruct} model to perform an information extraction task, categorizing each regretful comment accordingly. This model was first validated on our benchmark test set, where it achieved 92.08\% accuracy in extracting the correct reason 

\begin{figure}[htb]
    \centering
    \includegraphics[width=0.8\textwidth]{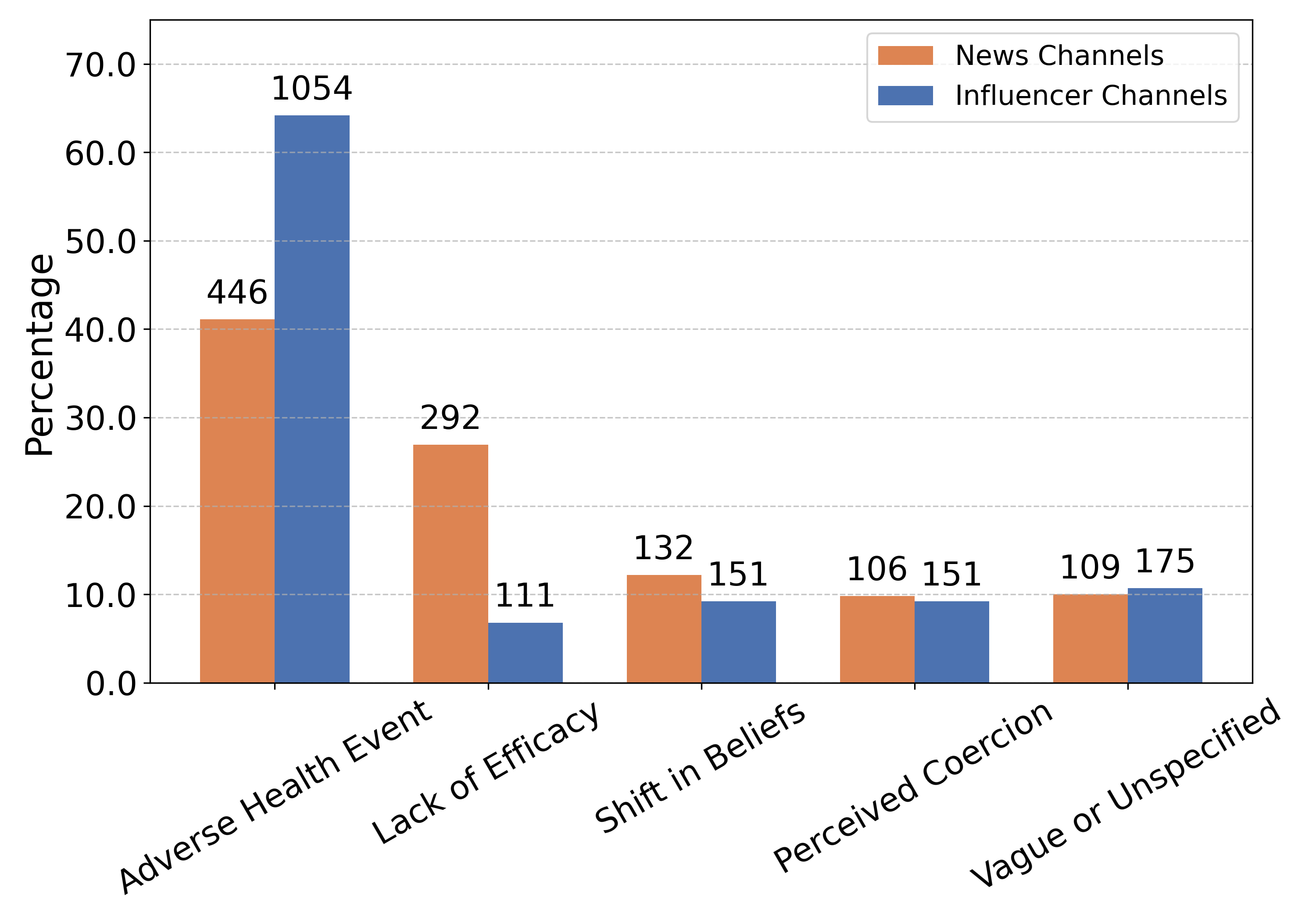}
    \caption{Distribution of Regret Reasons by Source Type. }
     \label{fig:regret_reasons_plot}
\end{figure}

Applying this validated method to the 2,727 comments expressing regret on $\mathcal{D}_\textit{wild}$, our analysis revealed that an \textit{Adverse Health Event} was the most common reason overall, accounting for 55.0\% of cases. However, the distribution of reasons differed significantly between news and influencer channels ($\chi^2$=248.84, p $<$ 0.001). As shown in Figure~\ref{fig:regret_reasons_plot}, influencer channels were dominated by discussions of adverse health events (64.2\%). In contrast, news channels featured a more balanced conversation where \textit{Lack of Efficacy} was a prominent secondary reason (26.9\%).


We also observed a significant difference in the reasons cited between first-person and vicarious accounts ($\chi^2$=45.94, p $<$ 0.001). While \textit{Adverse Health Event} was the most common reason for both groups, it was more dominant in vicarious narratives (61.3\%) than in first-person ones (52.1\%). Conversely,  \textit{Perceived Coercion} was more than twice as likely to be cited as a reason for regret in first-person accounts (11.6\%) compared to vicarious accounts (4.9\%). This supports previous research identifying perceived coercion as a key driver of regret~\cite{tayhan2025nursing} and highlights that the feeling of diminished autonomy is a powerful and personal component of this sentiment 

\subsubsection{Perceptual Differences in Annotation:}
We observed moderate inter-rater agreement among our politically diverse annotators across all three annotation tasks, as detailed in Table~\ref{tab:annotator_agreement}. Overall agreement (Fleiss' Kappa) was highest for the Subject task ($\kappa=0.5089$) and slightly lower for Regret ($\kappa=0.4480$) and Vaccinated ($\kappa=0.4272$). Our observed agreement aligns with prior literature~\cite{weerasooriya2023vicarious,crowl2025measuring}.

A Chi-square test revealed that political affiliation significantly influenced the subjective task of classifying a comment's \textit{Subject} ($\chi^2$=22.85, p $<$ 0.001). Conversely, we found no systematic impact on the more factual \textit{Vaccinated} or sentiment-based \textit{Regret} judgments, as detailed in Table~\ref{tab:systematic_differences}.


\begin{table}[htbp]
\centering
\small
\resizebox{\linewidth}{!}{%
\begin{tabular}{lcccc}
\toprule
\textbf{Task} & \textbf{Fleiss' $\kappa$} & \textbf{Dem vs Rep $\kappa$} & \textbf{Dem vs Ind $\kappa$} & \textbf{Rep vs Ind $\kappa$} \\
\midrule
Subject & 0.5089 & 0.5093 & 0.5140 & 0.5046 \\
Vaccinated & 0.4272 & 0.4380 & 0.4286 & 0.4330 \\
Regret & 0.4480 & 0.4246 & 0.4967 & 0.3677 \\
\bottomrule
\end{tabular}%
}
\caption{Inter-Annotator Agreement by Task}
\label{tab:annotator_agreement}
\end{table}

\begin{table}[htbp]
\centering
\small
\begin{tabular}{lcc}
\toprule
\textbf{Task} & \textbf{Chi-square ($\chi^2$)} & \textbf{p-value} \\
\midrule
Subject & 22.8515 & 0.0001 \\
Vaccinated & 1.6875 & 0.4301 \\
Regret & 3.1468 & 0.2073 \\
\bottomrule
\end{tabular}
\caption{Systematic Differences in Annotation by Political Affiliation}
\label{tab:systematic_differences}
\end{table}

\subsubsection{Model Alignment with Annotator Politics:}
On the subset of 398 comments where annotators disagreed on the Regret label (always in a 2-vs-1 split), we analyzed our pipeline's predictions to check for political alignment. The model's final output sided with the majority opinion 70.1\% of the time. The model's alignment with Democratic (55.8\%), Republican (53.3\%), and Independent (61.1\%) annotators was not statistically different ($\chi^2$ = 2.22, p = 0.3298). This suggests that for this task, the model does not systematically favor the perspective of one political group over the others.

\subsubsection{Temporal Study of Regret:}
\begin{figure}[htb]
    \centering
    \includegraphics[width=0.8\textwidth]{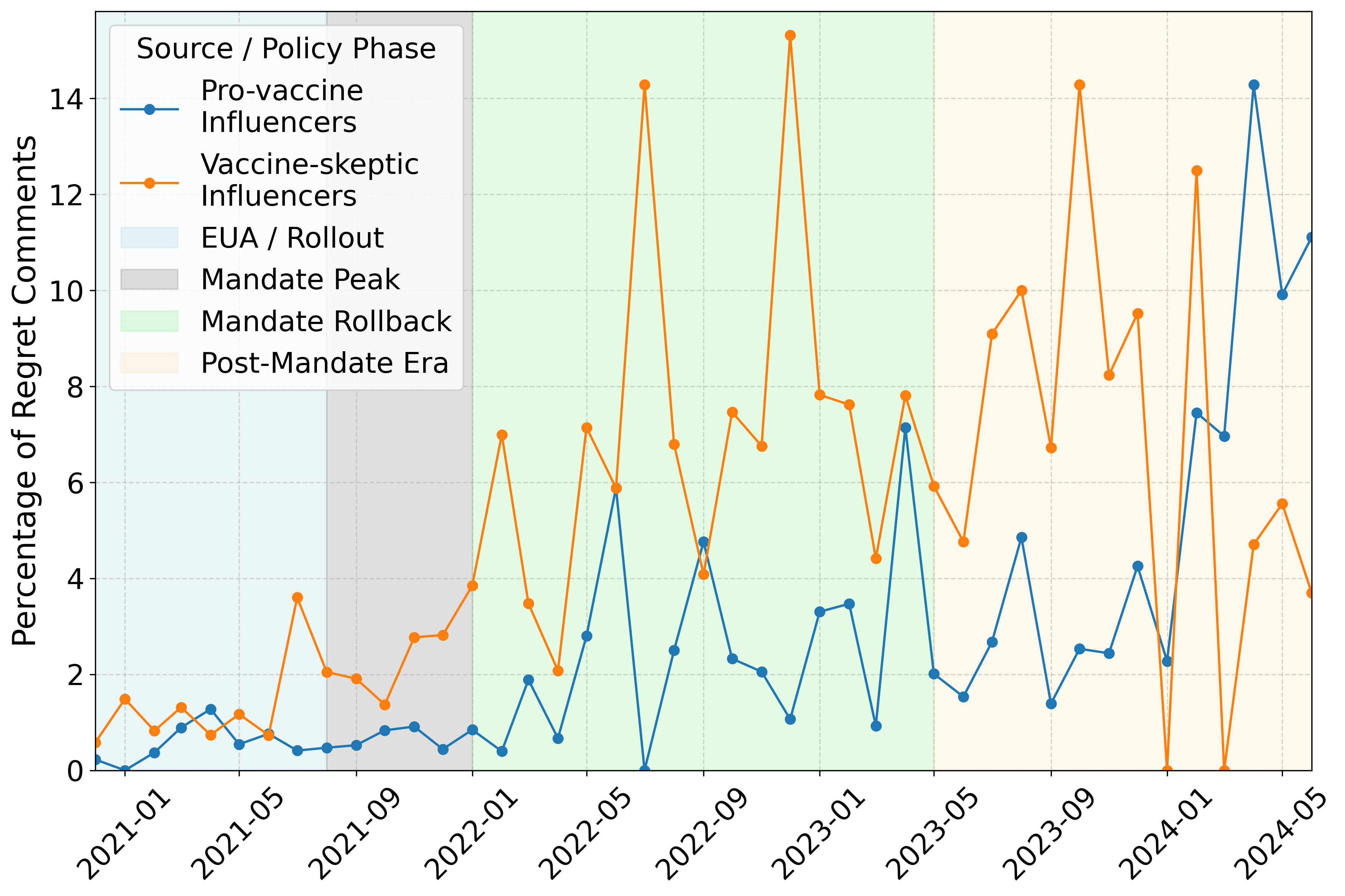}
    \caption{Temporal Distribution of Regret Comments over time across zones}
    \label{fig:regret_over_time_influencers}
\end{figure}

Figure~\ref{fig:regret_over_time_influencers} shows the temporal distribution of vaccine-related regret comments expressed by pro-vaccine and vaccine-skeptic influencers from 2020 through mid-2024, overlaid with key U.S. COVID vaccine policy phases. The graph is divided into four zones: Zone A (EUA \& Rollout) marks the earliest phase of public awareness, characterized by high uncertainty and the emergence of early regret commentary; Zone B (Mandate Peak) corresponds to the introduction of federal employee and contractor mandates, a period associated with heightened polarization and visible spikes in regret, particularly among vaccine-skeptic influencers; Zone C (Rollback) captures the easing of legal enforcement and restrictions, potentially correlating with a decline in overt regret discourse; and Zone D (Post-Mandate Era) reflects a stabilized regulatory environment where vaccine sentiment is increasingly shaped by individual and localized narratives. Throughout these phases, vaccine-skeptic influencers exhibit a rising trajectory of regret commentary, peaking dramatically by the end, while pro-vaccine influencers maintain comparatively lower and more sporadic levels of regret expression. While on the other end we do not see any noticeable differences in trend for the news channels.


\subsubsection{Qualitative Error Analysis:}
To better understand the model's performance, we conducted a qualitative analysis of misclassified examples in $\mathcal{D}_\textit{test}$. This revealed that the lower F1-score for subject classification is an expected artifact of our pipeline's design, which prioritizes efficiency. The initial NLI relevance filter effectively discards large volumes of irrelevant comments, which substantially improves processing speed and the stability of the downstream LLM. This efficiency creates an occasional mismatch with human annotations on irrelevant comments, artificially lowering the performance metric for the \textit{unspecified} subject class. This is an accepted trade-off, however, as it does not impact the accuracy of the primary regret classification task. 


\subsection{Limitations}
While our study offers significant insights to public's vaccine stance, it has some limitations that must be considered when interpreting the results.

First, our dataset of YouTube comments, like the official Vaccine Adverse Event Reporting System (VAERS) \cite{shimabukuro2015safety} rely on spontaneous, voluntary reports subject to significant self-selection bias; therefore, our findings cannot be used to infer causality or calculate the true incidence rate of vaccine regret or adverse events in the general population. Second, while our pipeline performed well, the classification of a nuanced human emotion like regret is inherently challenging. Sarcasm and complex expressions can still be misinterpreted by advanced LLMs~\cite{crowl2025measuring}, and the F1-score of 0.83 for the regret class indicates some degree of model error is unavoidable. Finally, our analysis is limited to English-language comments on a single platform (YouTube). The dynamics of post-vaccination sentiment may differ significantly in other languages, cultural contexts, and on other social media platforms.




\subsection{Implications and Future Work}

Our research has several implications. For public health officials, our work provides a scalable method for monitoring public sentiment and identifying the primary drivers of vaccine regret. Understanding that narratives of coercion, adverse events, and lack of efficacy are central to this sentiment can help refine public health communication and improve 
Proactively addressing public concerns about adverse events and communicating the rationale behind public health mandates are also essential to maintain institutional trust \cite{souvatzi2024trust}. The stark differences between mainstream news and influencer communities underscore the need for tailored outreach strategies that address the specific concerns circulating in different online ecosystems.


Future research directions include expanding this analysis to other social media platforms, languages, and cultural contexts to create a more holistic picture of global post-vaccination sentiment. A longitudinal study tracking how these regret narratives evolve over time, particularly in response to new public health developments, would be highly valuable. While our human annotation pipeline accounts for political diversity, extending the annotation pipeline to vicarious interactions~\cite{weerasooriya2023vicarious} can help us understand how well raters can represent out-group values and opinions on politically sensitive topics. 
Finally, refining the classification of regret reasons, for instance by distinguishing between mild and severe adverse events, could provide even more granular insights for public health intervention.

\section{Conclusion}
In this study, we introduced a novel computational framework to quantify and analyze the phenomenon of \textit{vaccine buyer's remorse} within a large corpus of YouTube comments. Our findings reveal that while expressions of regret are a persistent feature of online discourse, they represent a small fraction of the overall conversation. This sentiment is most prevalent on influencer channels and is primarily driven by narratives of adverse health events, perceived lack of efficacy, and feelings of coercion. By differentiating between first-person and vicarious reports and identifying primary drivers of regret, such as adverse health events and perceived lack of efficacy, this study provides key insights into post-vaccination attitudes. In the end, this approach provides a scalable method for public health monitoring and highlights the need for targeted communication efforts that address specific public worries to build trust in a growingly divided and polarized information environment.

\section*{Acknowledgments}
Miles Stanley and Ashiqur R. KhudaBukhsh were partially supported by NSF Award \#2447631.

\bibliographystyle{abbrv}
\bibliography{references.bib}

\begingroup 
\setcounter{secnumdepth}{1}
\section{Supplementary Information}
\appendix
\section{Data Collection and Filtering Details}
\label{app:data_collection}

\subsection{Full List of YouTube Channels}
\label{app:channels}
Table \ref{tab:data_summary} and Table \ref{tab:influencer_data} list all the mainstream news and influencer YouTube channels from which comments were collected for this study. For each influencer we performed a keyword search with the same keywords from \ref{app:keywords}: Initial Vaccine-Relevant Filtering to find relevant videos. 6 videos found in this search were private and we were unable to gather comments for them.

\begin{table*}[t]
    \centering
    \caption{Summary of the unfiltered data collected from news organization channels.}
    \label{tab:data_summary}
    \begin{tabular}{lccc}
        \toprule
        \textbf{News Channel} & \textbf{\# of Videos} & \textbf{\# of Comments} \\
        \midrule
        Fox News & 23902 & 31340801 \\
        CNN & 12490 & 22139179  \\
        MSNBC & 29494 & 26827950 \\
        \midrule
        \textbf{Total} & \textbf{65886} & \textbf{80307930} \\
        \bottomrule
    \end{tabular}
\end{table*}

\begin{table*}[t]
    \centering
    \caption{Summary of data collected from individual influencers, grouped by vaccine opinion.}
    \label{tab:influencer_data}
    \begin{tabular}{llcc}
        \toprule
        \textbf{Vaccine Opinion} & \textbf{Influencer} & \textbf{\# of Videos} & \textbf{\# of Comments} \\
        \midrule
        
        Pro-vaccine & David Pakman & 412 & 204,410 \\
        & The Young Turks & 177 & 151,068 \\
        & Secular Talk & 88 & 57,175 \\
        & Hasan Piker & 14 & 10,961 \\
        & Destiny & 10 & 4,915 \\
        \cmidrule(l){2-4}
        & \textbf{Pro-vaccine Subtotal} & \textbf{701} & \textbf{428,529} \\
        \midrule
        
        Vaccine-skeptic & Tim Pool & 32 & 129,901 \\
        & Joe Rogan & 9 & 126,414 \\
        & Megyn Kelly & 88 & 51,520 \\
        & The Rubin Report & 39 & 45,364 \\
        & Ben Shapiro & 25 & 41,942 \\
        & Turning Point USA & 82 & 13,176 \\
        & PragerU & 4 & 9,256 \\
        & Steven Crowder & 1 & 1,600 \\
        \cmidrule(l){2-4}
        & \textbf{Vaccine-skeptic Subtotal} & \textbf{280} & \textbf{419,173} \\
        \midrule
        
        \textbf{Total} & & \textbf{981} & \textbf{847,702} \\
        \bottomrule
    \end{tabular}
\end{table*}

\subsection{Keyword Lists for Filtering}
\label{app:keywords}
\begin{itemize}
    \item \textbf{Initial Vaccine-Relevant Filtering:} \texttt{'vaccine', 'vaccination', 'vaccinated', 'vaccines', 'pfizer', 'moderna', 'johnson', 'j\&j', 'booster', 'shot', 'dose', 'jab', 'mrna', 'immunization', 'vaxx', 'antivaxx', 'anti-vaxx'}
    \item \textbf{Keywords for Regret-Enrichment Sampling:} \texttt{r"wishes? \textbackslash w+ (never|hadn't)",
    r"ruined (my|her|his|their) (life|health|body)",
    r"after (the|my|her|his|their) (vaccine|first|second|third|booster) (dose)?",
    r"destroyed (my|her|his|their) (life|health|body)",
    r"(i|he|she|they) (was|were) injured by the vaccine",
    r"(i|he|she|they|we) gave in to (the )?pressure",
    r"was perfectly healthy before",
    r"ever since (i|he|she|they) got (it|(the )? (shot|jab|vaccine))",
    r"should have said no",
    r"the vaccine caused",
    r"what have i done"}
\end{itemize}

\section{Annotation Study Materials}
\label{app:annotation_materials}
\subsection{Full Annotator Instructions}
\label{app:annotator_instructions}
\textit{The following instructions were provided to all annotators participating in the benchmark dataset creation.}

\paragraph{1. Goal} The purpose of this project is to carefully read user comments about vaccines and classify them based on three key pieces of information: who the comment is about, their vaccination status, and their feelings about their decision.

\paragraph{2. The Annotation Task} For each comment you are shown, you will answer a series of up to three questions. Please note that some questions will only appear based on your answer to the previous question.

\subparagraph{Question 1: Who is the subject of the comment?}
This question asks you to identify the main person or group being discussed in the comment.
\begin{itemize}
    \item \textbf{self:} The author of the comment is talking about their own personal experience.
        \begin{itemize}
            \item \textit{Example:} "I got the shot and I feel fine."
            \item \textit{Example:} "I regret my decision to get vaccinated."
        \end{itemize}
    \item \textbf{other:} The author is talking about a specific person or a small, well-defined group of people they know (e.g., a friend, family member, celebrity, "my friends," "my whole family").
        \begin{itemize}
            \item \textit{Example:} "My daughter had a bad reaction."
            \item \textit{Example:} "My uncle wishes he never got it."
        \end{itemize}
    \item \textbf{unspecified:} The comment is off-topic, confusing, it's impossible to tell who the subject is, or it refers to a broad, non-specific group of people (e.g., "people," "everyone," "those who got the shot").
        \begin{itemize}
            \item \textit{Example:} "This is all about politics and has nothing to do with health."
            \item \textit{Example:} "Everyone who got the vaccine was lied to."
        \end{itemize}
\end{itemize}

\subparagraph{Question 2: Is the subject vaccinated?}
This question will only appear if you selected \textit{self} or \textit{other} in Question 1.
\begin{itemize}
    \item \textbf{Yes, they were vaccinated:} The comment clearly states or strongly implies the subject received a vaccine.
        \begin{itemize}
            \item \textit{Example:} "I got my second dose last week."
            \item \textit{Example:} "She had a bad reaction to the jab."
        \end{itemize}
    \item \textbf{No, or it's unclear:} The comment clearly states the subject did not receive a vaccine or it does not provide enough information to be certain.
        \begin{itemize}
            \item \textit{Example (No):} "I'm not vaccinated and I stand by my choice."
            \item \textit{Example (Unclear):} "My dad is worried about the mandates."
        \end{itemize}
\end{itemize}

\subparagraph{Question 3: Does the subject express regret?}
This question will only appear if you answered \textit{Yes, they were vaccinated} in Question 2. The Key: You are trying to capture the subject's feeling about their decision, not the commenter's opinion.
\begin{itemize}
    \item \textbf{Yes, they express regret:} Choose this only if the comment indicates the vaccinated subject feels negative about their decision. This can be:
        \begin{itemize}
            \item An explicit statement: "I regret it," "I wish I never got it," "It was a mistake."
            \item A strong implication, such as when the comment describes severe, life-altering negative health outcomes that the subject directly attributes to the vaccine, or when the subject actively warns others not to get the vaccine because of their own negative experience.
            \item \textit{Example:} "My friend has had constant heart problems ever since her second shot."
            \item \textit{Example:} "Don't get the vaccine, ever since I got it I have felt tired and had rashes."
        \end{itemize}
    \item \textbf{No, or it's unclear:} Choose this for all other cases where a specific person was vaccinated. This includes:
        \begin{itemize}
            \item Positive or neutral statements: "I'm glad I got it," "I got my booster shot today."
            \item Minor side effects framed as temporary or expected: "The headache was no joke, but I'm happy to be protected."
            \item General anger from the commenter that is not attributed to the subject: "My uncle got the shot and felt sick. I think this whole thing is poison!" (Here, the commenter is angry, but the uncle's feeling of regret is not stated. You would choose "No, or it's unclear").
        \end{itemize}
\end{itemize}

\paragraph{3. Final Reminders}
\begin{itemize}
    \item \textbf{Prioritization:} If a comment mentions multiple subjects, prioritize the subject who was vaccinated and expresses regret.
    \item \textbf{Uncertainty:} When in doubt, choose the "unspecified" or "unclear" option. It is better to choose an unclear option than to guess an incorrect label.
    \item \textbf{Focus on Text:} Base your judgment only on the text provided. Do not make assumptions or use outside knowledge.
\end{itemize}

\section{Model Performance}
\label{app:all_model_performance}

\begin{table}[htb]
\centering
\small
\setlength{\tabcolsep}{2pt}
\begin{tabular}{llcccc}
\hline
\textbf{Task} & \textbf{Metric/Class} & \textbf{Precision} & \textbf{Recall} & \textbf{F1} & \textbf{Support} \\
\hline
\multirow{5}{*}{\textbf{Subject}} & Accuracy: 81.50\% & & & & \\
& other & 0.86 & 0.74 & 0.80 & 181 \\
& self & 0.91 & 0.94 & 0.93 & 139 \\
& unspecified & 0.60 & 0.76 & 0.67 & 80 \\
& \textit{macro avg} & \textit{0.79} & \textit{0.82} & \textit{0.80} & \textit{400} \\
\hline
\multirow{4}{*}{\textbf{Vaccinated}} & Accuracy: 87.25\% & & & & \\
& 0 (No/Unspecified) & 0.85 & 0.87 & 0.86 & 181 \\
& 1 (Yes) & 0.89 & 0.88 & 0.88 & 219 \\
& \textit{macro avg} & \textit{0.87} & \textit{0.87} & \textit{0.87} & \textit{400} \\
\hline
\multirow{4}{*}{\textbf{Regret}} & Accuracy: 86.00\% & & & & \\
& 0 (No) & 0.91 & 0.90 & 0.91 & 299 \\
& 1 (Yes) & 0.71 & 0.75 & 0.73 & 101 \\
& \textit{macro avg} & \textit{0.81} & \textit{0.82} & \textit{0.82} & \textit{400} \\
\hline
\end{tabular}
\caption{Performance Metrics of the Classification Pipeline}
\label{tab:model_performance_pipeline}
\end{table}

\begin{table}[htbp]
\centering
\small
\setlength{\tabcolsep}{4pt}
\caption{Performance of Zero-Shot Model for Relationship Classification}
\label{tab:relationship_performance}
\begin{tabular}{lcccc}
\hline
\textbf{Category} & \textbf{Precision} & \textbf{Recall} & \textbf{F1-Score} & \textbf{Support} \\
\hline
Spouse or Partner & 1.00 & 0.90 & 0.95 & 10 \\
Family Member & 1.00 & 1.00 & 1.00 & 32 \\
Friend & 1.00 & 0.91 & 0.95 & 11 \\
Health Care Provider & 1.00 & 0.50 & 0.67 & 2 \\
Public Figure & 1.00 & 0.83 & 0.91 & 24 \\
Other Acquaintance & 0.69 & 0.95 & 0.80 & 19 \\
Unspecified & 0.88 & 0.88 & 0.88 & 42 \\
\hline
\textbf{Weighted Avg} & \textbf{0.92} & \textbf{0.91} & \textbf{0.91} & \textbf{140} \\
\hline
\end{tabular}
\end{table}

\begin{table}[htbp]
\centering
\small
\setlength{\tabcolsep}{4pt}
\caption{Performance of Zero-Shot Model for Reason for Regret Extraction }
\label{tab:reasons_performance}
\begin{tabular}{lcccc}
\hline
\textbf{Category} & \textbf{Precision} & \textbf{Recall} & \textbf{F1-Score} & \textbf{Support} \\
\hline
Adverse Health Event & 0.96 & 0.93 & 0.95 & 58 \\
Perceived Coercion & 1.00 & 0.85 & 0.92 & 13 \\
Lack of Efficacy & 0.94 & 0.89 & 0.91 & 18 \\
Shift in Beliefs & 0.67 & 1.00 & 0.80 & 6 \\
Vague or Unspecified & 0.75 & 1.00 & 0.86 & 6 \\
\hline
\textbf{Weighted Avg} & \textbf{0.93} & \textbf{0.92} & \textbf{0.92} & \textbf{101} \\
\hline
\end{tabular}
\end{table}

\section{Reasons for Regret by Narrative Perspective}
\label{app:reasons_by_perspective}

\begin{figure}[htb]
    \centering
    \hspace{-20pt}
    \includegraphics[width=0.5\textwidth]{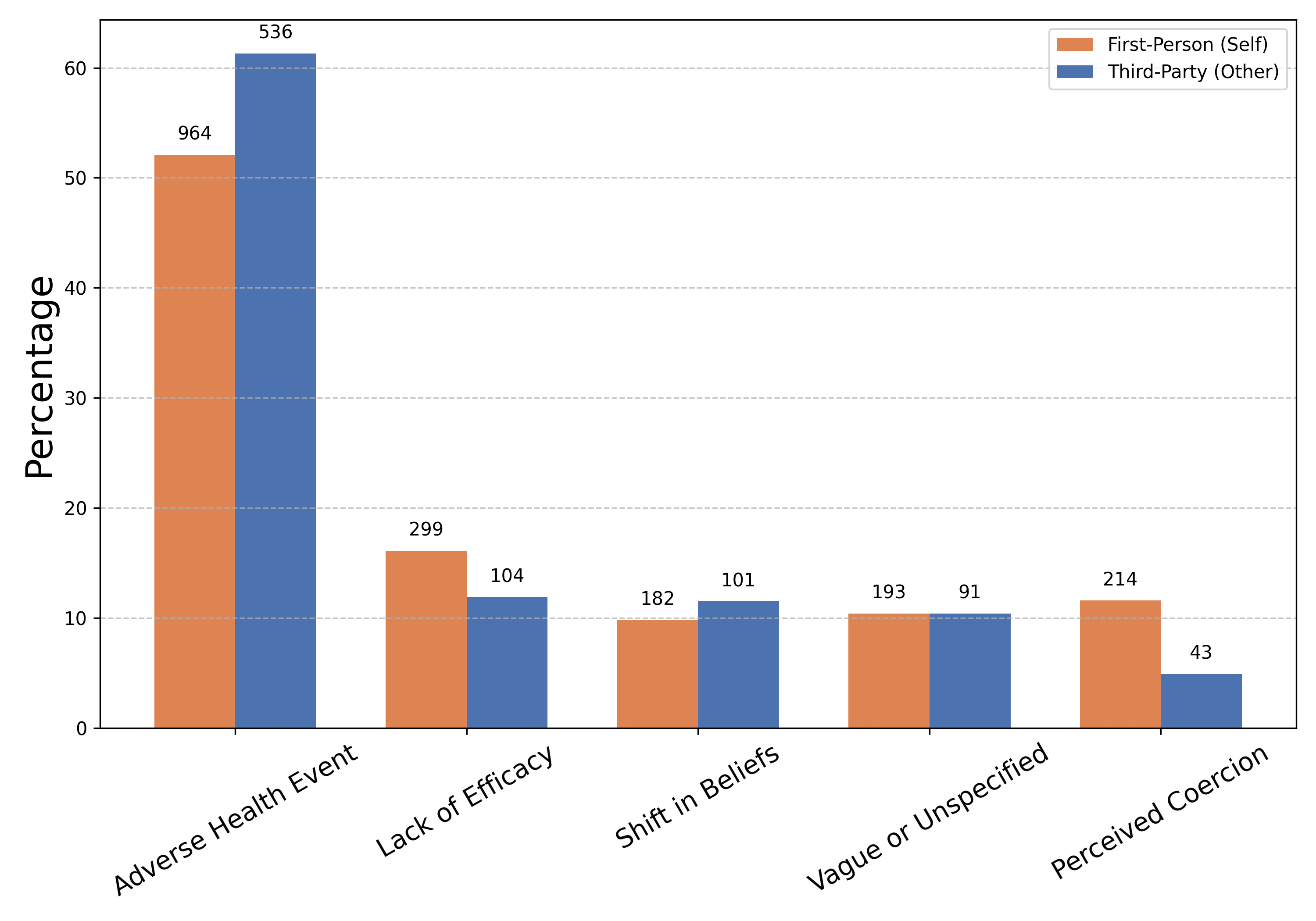}
    \caption{Distribution of Regret Reasons by Narrative Perspective.}
    \label{fig:regret_narrative}
\end{figure}

\section{Model Prompts and Hyperparameters}
\label{app:prompts_hyperparams}

\begin{figure}
\begin{lstlisting}[style=promptstyle]
You are a data analyst. Analyze the user's vaccine comment and provide a single JSON object with the keys: `subject`, `vaccinated`, and `regret`.

**Core Rules:**
1. **Regret Scope:** `regret` = 1 may ONLY be assigned to a specific subject (`"self"` or `"other"`) that has taken the vaccine (`vaccinated` = 1). For general statements about groups, regret MUST be 0.
2. **Regret Perspective:** The regret must be the subject's own reported feeling, not the commenter's projected opinion.
3. **Vaccination Status:** `vaccinated` = 1 may ONLY be assigned to a specific subject (`"self"` or `"other"`) that has taken the vaccine.

**JSON Schema & Values:**
- `subject`: "self" | "other" | "unspecified"
- `vaccinated`: 0 | 1
- `regret`: 0 | 1

**Regret Definitions:**
- 1: A specific subject (`self`/`other`) explicitly states regret, warns others based on their outcome, or describes severe negative health results from the vaccine.
- 0: The subject doesn't express regret or the subject is a general group without personal sentiment or a negative sentiment is only from the commenter.

Now, analyze the following user comment and provide only the JSON output without any commentary.
**Comment:** {comment}
\end{lstlisting}
\caption{Prompt used for the Llama-3.1 "Expert Reasoner" model and zero-shot prompting.}
\label{fig:prompt_expert_reasoner}
\end{figure}

\begin{figure}
\begin{lstlisting}[style=promptstyle]
You are a data analyst. Analyze the user's vaccine comment and provide a single JSON object with the key: `regret`.

**Core Rules:**
1.  **Regret Scope:** `regret` = 1 may ONLY be assigned to a specific subject that has taken the vaccine. For general statements about groups, regret MUST be 0.
2.  **Regret Perspective:** The regret must be the subject's own reported feeling, not the commenter's projected opinion.

**JSON Schema:**
-   `regret`: 0 | 1

**Regret Definitions:**
-   1: A specific subject explicitly states regret, warns others based on their outcome, or describes severe negative health results from the vaccine.
-   0: The subject doesn't express regret or the subject is a general group without personal sentiment or the negative sentiment is only from the commenter.

Now, analyze the following user comment and provide only the JSON output without any commentary.
**Comment:** {comment}
\end{lstlisting}
\caption{Prompt used for the political group-specific models}
\label{fig:prompt_political_models}
\end{figure}

\begin{figure}
\begin{lstlisting}[style=promptstyle]
You are a data analyst. Analyze the user's vaccine comment and provide a single JSON object with the keys: `subject`, `vaccinated`, and `regret`.

**Core Rules:**
1.  **Regret Scope:** `regret` = 1 may ONLY be assigned to a specific subject (`\"self\"` or `\"other\"`) that has taken the vaccine (`vaccinated` = 1). For general statements about groups, regret MUST be 0.
2.  **Regret Perspective:** The regret must be the subject's own reported feeling, not the commenter's projected opinion.
3.  **Vaccination Status:** `vaccinated` = 1 may ONLY be assigned to a specific subject (`\"self\"` or `\"other\"`) that has taken the vaccine.

**JSON Schema & Values:**
-   `subject`: "self" | "other" | "unspecified"
-   `vaccinated`: 0 | 1
-   `regret`: 0 | 1

**Regret Definitions:**
-   1: A specific subject (`self`/`other`) explicitly states regret, warns others based on their outcome, or describes severe negative health results from the vaccine.
-   0: The subject doesn't express regret or the subject is a general group without personal sentiment or the negative sentiment is only from the commenter.

**Examples:**
1.  **Comment:** "my friend regrets getting the vaccine. her period every 2 weeks isn't normal."
    **JSON:** {"subject": "other", "vaccinated": 1, "regret": 1}

2.  **Comment:** "My brother got his second dose last week."
    **JSON:** {"subject": "other", "vaccinated": 1, "regret": 0}

3.  **Comment:** "don't get the vaccine, ever since i got it i have felt tired and had rashes."
    **JSON:** {"subject": "self", "vaccinated": 1, "regret": 1}

4.  **Comment:** "I wonder if the people getting the vaccine will regret it in 10 years when they have serious side effects"
    **JSON:** {"subject": "unspecified", "vaccinated": 0, "regret": 0}

5.  **Comment:** "cnn should interview the people who got the vaccine and regret it after serious side effect"
    **JSON:** {"subject": "unspecified", "vaccinated": 0, "regret": 0}

Now, analyze the following user comment and provide only the JSON output without any commentary.
**Comment:** {comment}
\end{lstlisting}
\caption{Few shot prompt}
\label{fig:prompt_few_shot}
\end{figure}

\begin{figure}
\begin{lstlisting}[style=promptstyle]
You are an expert qualitative data analyst. Your task is to analyze a user comment and identify the **primary reason** for their expressed regret about a vaccination decision.

You will be given a comment where the user has already been identified as expressing regret. Your sole function is to classify the reason for that regret into one of the following five categories.

**CRITICAL INSTRUCTIONS:**
1. Your output MUST be ONLY a single, raw JSON object.
2. The JSON object must contain a single key: `"reason_for_regret"`.
3. The value for this key must be ONE of the following exact strings:
    * `"Adverse_Health_Event"`
    * `"Perceived_Coercion"`
    * `"Lack_of_Efficacy"`
    * `"Shift_in_Beliefs"`
    * `"Vague_or_Unspecified"`

---

**CATEGORY DEFINITIONS & RULES:**

* **`"Adverse_Health_Event"`**
    * Assign this category if the regret is linked to any negative physical health outcome.
    * **Look for:** Mentions of specific side effects (fever, pain, fatigue), new or worsening chronic conditions (heart problems, autoimmune issues, blood clots), or reproductive health concerns (menstrual changes, fertility worries).
    * *Example signals:* "I've had constant heart problems ever since...", "The side effects were unbearable...", "My life is ruined by this new condition..."

* **`"Perceived_Coercion"`**
    * Assign this category if the regret stems from feeling forced, pressured, or mandated to get the vaccine, regardless of health outcomes. The core emotion is a loss of autonomy or choice.
    * **Look for:** Mentions of work, school, or travel requirements.
    * *Example signals:* "...only got it to keep my job.", "I wish I had stood my ground against the mandate.", "I regret giving in to the pressure."

* **`"Lack_of_Efficacy"`**
    * Assign this category if the regret is based on the belief that the vaccine did not work as promised. The focus is on performance, not safety.
    * **Look for:** Statements about getting infected despite being vaccinated or the vaccine not stopping transmission.
    * *Example signals:* "I regret it because I still got Covid twice.", "It was useless since it didn't stop the spread."

* **`"Shift_in_Beliefs"`**
    * Assign this category if the regret is due to a change of mind based on new information or research the user encountered *after* getting vaccinated.
    * **Look for:** Phrases that indicate a change in perspective over time.
    * *Example signals:* "Knowing what we know now, I wish I hadn't.", "After doing my own research...", "I feel like we were lied to about the initial data."

* **`"Vague_or_Unspecified"`**
    * Assign this category ONLY if the user explicitly states regret but gives no specific reason from the categories above.
    * **Look for:** General statements of regret without any supporting cause.
    * *Example signals:* "It was the biggest mistake of my life.", "I regret it so much.", "Never again."

**PRIORITY RULE:** If a comment mentions multiple reasons (e.g., a side effect AND a mandate), choose the category that is most emphasized by the user as the primary source of their regret.

Now, classify the following comment.
\end{lstlisting}
\caption{Prompt used for the "Reasons for Regret" extraction task.}
\label{fig:prompt_reasons}
\end{figure}

\begin{figure}
\begin{lstlisting}[style=promptstyle]
You are a data analyst. Your task is to analyze a user comment and classify the relationship of the person experiencing vaccine regret to the author of the comment.

The comment provided is a vicarious account, meaning the author is telling someone else's story. Your sole function is to classify this relationship into one of the following categories.

**CRITICAL INSTRUCTIONS:**
1.  Your output MUST be ONLY a single, raw JSON object.
2.  The JSON object must contain a single key: `"relationship_to_author"`.
3.  The value for this key must be ONE of the following exact strings:
    * `"Spouse_or_Partner"`
    * `"Family_Member"`
    * `"Friend"`
    * `"Health_Care_Provider"`
    * `"Public_Figure"`
    * `"Other_Acquaintance"`
    * `"Unspecified"`

---

**CATEGORY DEFINITIONS & RULES:**

* **`"Spouse_or_Partner"`**
    * Assign this for a spouse or romantic partner.
    * **Look for:** "husband," "wife," "partner," "boyfriend," "girlfriend," "fiance."
    * *Example signal:* "My husband wishes he never got it."

* **`"Family_Member"`**
    * Assign this for any other familial relationship.
    * **Look for:** "mother," "father," "son," "daughter," "brother," "sister," "grandparent," "cousin," "uncle," "aunt."
    * *Example signal:* "My sister has had a terrible time..."

* **`"Friend"`**
    * Assign this category for peer friendships.
    * **Look for:** "friend," "best friend," "buddy."
    * *Example signal:* "My friend has been in er ever since his jab."

* **`"Health_Care_Provider"`**
    * Assign this category for medical professionals.
    * **Look for:** "my doctor," "our pediatrician," "a nurse I know."
    * *Example signal:* "My doctor has stopped recommending the vaccine and regrets getting it."

* **`"Public_Figure"`**
    * Assign this category for celebrities, politicians, athletes, or other well-known individuals.
    * **Look for:** Names of famous people.
    * *Example signal:* "I heard Eric Clapton regrets getting vaccinated."

* **`"Other_Acquaintance"`**
    * Assign this for more distant but still specific relationships.
    * **Look for:** "coworker," "colleague," "boss," "neighbor," "a guy from my church," "my instructor."
    * *Example signal:* "One of my coworkers has been out sick for weeks since the shot."

* **`"Unspecified"`**
    * Assign this category ONLY if the relationship is too vague to classify or is not stated.
    * **Look for:** "someone I know," "a person I met," or if the comment just launches into a story without defining the relationship.
    * *Example signal:* "I know someone who now has permanent heart damage."

Now, classify the relationship in the following comment.
**Comment:** {comment}
\end{lstlisting}
\caption{Prompt used for the "Relationship to Author" extraction task.}
\label{fig:prompt_relationship}
\end{figure}

\begin{table}[htbp]
\centering
\caption{Fine-tuning Hyperparameters for Llama-3.1 70b}
\label{tab:hyperparameters}
\begin{tabular}{lcc}
\toprule
\textbf{Hyperparameter} & Value\\
\midrule
Learning Rate & 2e-4\\
Learning Rate Scheduler & cosine \\
Batch Size & 1 \\
Gradient Accumulation Steps & 4\\ 
Number of Epochs & 3 \\
Weight Decay & 0.001 \\
Optimizer & paged\_adamw\_32bit \\
Warm-up Ratio & 0.03 \\
Max Gradient Norm & 0.3 \\
Quantization & 4-bit \\
LORA rank (r) & 16 \\
LORA alpha & 32 \\
\bottomrule
\end{tabular}
\end{table}

\section{Qualitative Examples}
\label{app:qualitative_examples}
\begin{table*}[htbp]
\centering
\caption{Examples of Pipeline Classifications}
\label{tab:qual_examples}
\begin{tabular}{p{0.5\textwidth}ccc}
\toprule
\textbf{Comment} & \textbf{Subject} & \textbf{Vaccinated} & \textbf{Regret} \\
\midrule
i am 69 years old.  i was vaccinated twice.  i am vaccine injured.  i cannot take the booster.  i have not been out in public since july of 2021 without a p100 respirator.  no one is going to tell me to get another risky vaxx.  no one! eric speaks for me. so all you high and mighty anti trumper's can stick it! & self & 1 & 1 \\
\midrule
don't believe anything don lemon says cnn lies you can pay anybody to say anything the right amount of money will make anybody say anything! best friends vaccinated mother has got a story to tell two and regrets ever taking this vaccine & other & 1 & 1 \\
\midrule
it hurts my heart knowing my great state of michigan is climbing in cases. the biggest b.s is that we the smart vaccinated ones if we end up in the hospital we aren't eligible for the infusion but unvaccinated people are. how is this even fair!?!? how can the non-vaccinated people be put above people who actually care for more then just themselves!? i don't like to talk ill about people but if your that ignorant not to get vaccinated and spread this crap because your scared the government is trying to spy on you. p.s you are already being tracked by your dang smart phones, smart cars and smart devices in your home duh! they don't  need to try and track you with a dang vaccine that is only trying g to get us back some what of normalcy! & self & 1 & 0 \\
\midrule
i wonder what's in the vaccine that they're pushing it so hard & unspecified & 0 & 0 \\
\bottomrule
\end{tabular}
\end{table*}

\begin{table*}[htbp]
\centering
\caption{Examples of Comments with Annotator Disagreement for Regret Classification}
\label{tab:disagreement_examples}
\begin{tabular}{p{0.4\textwidth}cccc}
\toprule
\textbf{Comment} & \textbf{Dem.} & \textbf{Rep.} & \textbf{Ind.} & \textbf{Pipeline} \\
\midrule
{i caught covid wit da vaccine} & 1 & 0 & 0 & 0 \\
\midrule
{if people don't want the vaccine, then they don't have to get it. two weeks after my second dose, i had to be rushed to the emergency room. i was diagnosed with a heart condition called sinus tachycardia. im only 27, in good health, and have no family history of any heart conditions whatsoever. so i know why people are iffy about the vaccine, we do not know the long term effects} & 0 & 0 & 1 & 1 \\
\bottomrule
\end{tabular}
\end{table*}

\begin{table}[htbp]
\centering
\small
\setlength{\tabcolsep}{2pt}
\caption{Distribution of Vaccine Regret Sentiment Across Source Categories}
\label{tab:regret_distribution}
\begin{tabular}{lccc}
\hline
\textbf{Source Category} & \textbf{Relevant Comments} & \textbf{Regret Rate} \\
\hline
Fox News & 56,822 & 0.7\% \\
CNN & 57,883 & 0.7\% \\
MSNBC & 43,968 & 0.6\% \\
\hline
Pro-Vaccine Influencers & 43,527 & 1.0\% \\
Vaccine-Skeptic Influencers & 41,347 & 2.9\% \\
\hline
\textbf{Total} & \textbf{243,547} & \textbf{1.1\%} \\
\hline
\end{tabular}
\end{table}

\begin{table}[htbp]
\centering
\small
\setlength{\tabcolsep}{2pt}
\caption{Results of NLI Relevance Filtering by Source Category}
\label{tab:relevance_filtering}
\begin{tabular}{lccc}
\hline
\textbf{Source Category} & \textbf{Total Comments} & \textbf{Relevance Rate} \\
\hline
Fox News & 100,000 & 56.8\% \\
CNN & 100,000 & 58.9\% \\
MSNBC & 100,000 & 44.0\% \\
\hline
\textit{Subtotal News} & \textit{300,000} & \textit{52.9\%} \\
\hline
Pro-Vaccine Influencers & 150,000 & 29.0\% \\
Vaccine-Skeptic Influencers & 150,000 & 27.6\% \\
\hline
\textit{Subtotal Influencers} & \textit{150,000} & \textit{28.3\%} \\
\hline
\textbf{Total} & \textbf{600,000} & \textbf{40.6\%} \\
\hline
\end{tabular}
\end{table}

\section{Error Analysis}
\label{app:error_analysis}
This appendix provides a detailed qualitative analysis of the misclassifications made by our two-stage hybrid inference pipeline. By examining specific error patterns, we aim to offer deeper insight into the pipeline's performance, its limitations, and the inherent challenges of classifying nuanced, user-generated content. Errors can originate from one of two points in our pipeline, the NLI Relevance Filter or the Llama-3.1 "Expert Reasoner."

\subsection{Errors from the NLI Relevance Filter}
As mentioned in the results section, the NLI Relevance Filter's design can create an apparent mismatch in subject classification. This happens when the filter correctly discards comments irrelevant to vaccines, which the pipeline then defaults to the unspecified subject label. Human annotators, however, sometimes assigned a self or other label to these same comments, which were often "hard negatives" included to test the system's robustness.

Our analysis of the 400-comment test set confirms this: every comment filtered out by the NLI model was indeed not relevant to vaccines (e.g., referring to "shot" in the context of firing a weapon). Table~\ref{tab:nli_error_examples} contains examples of these subject misclassifications caused by the NLI model (a relevance of 0 corresponds to a label of irrelevant).

\begin{table*}[htbp]
\centering
\setlength{\tabcolsep}{2pt}
\caption{Examples of NLI Errors}
\label{tab:nli_error_examples}
\begin{tabular}{|p{0.4\textwidth}|ccc|cccc|}
\hline
\multirow{2}{*}{\textbf{Comment Text}} & \multicolumn{3}{c|}{\textbf{True Label}} & \multicolumn{4}{c}{\textbf{Predicted Label}} \\
& \textbf{Subject} & \textbf{Vaccinated} & \textbf{Regret} & \textbf{Subject} & \textbf{Vaccinated} & \textbf{Regret} & \textbf{Relevance} \\
\hline
"not a big deal"- joe biden after the first balloon was shot down." & other & 0 & 0 & unspecified & 0 & 0 & 0 \\
\hline
hey johnson, here is religion for you: “and the lord god formed man of the dust of the ground, and breathed into his nostrils the breath of life; and man became a living soul.” - genesis 2:7  so according to your bible, a baby is only a living soul after the first breath." & other & 0 & 0 & unspecified & 0 & 0 & 0 \\
\hline
we can be sure dawn bancroft is only one of many of the rioters at the capitol on jan. 6 that wishes they hadn't shot video of themselves responding to djt's incitement to violence. if all criminals were as stupid as she is, we wouldn't need law enforcement any more. just round up the cell phones, watch the videos, and throw them in jail. & other & 0 & 0 & unspecified & 0 & 0 & 0 \\
\hline
\end{tabular}
\end{table*}

\subsection{Errors from the Llama-3.1 "Expert Reasoner"}
The most common LLM errors involved the misclassification of regret, a task made challenging by the nuanced ways people express dissatisfaction. As shown in Tables~\ref{tab:false_positive_examples} and \ref{tab:false_negative_examples}, the model struggles with both false positives and false negatives. It incorrectly infers regret when it mistakes strong negative sentiment, such as anger or frustration, for personal regret, or when it misinterprets a commenter's projection of regret onto another person ("i bet he regrets"). Conversely, the model fails to detect regret when the sentiment is implied rather than explicitly stated. It overlooks strong contextual clues like feelings of coercion, severe adverse outcomes, or sarcasm.

\begin{table*}[htbp]
\centering
\caption{Examples of False Positives (Regret)}
\label{tab:false_positive_examples}
\begin{tabular}{|p{0.4\textwidth}|ccc|ccc|}
\hline
\multirow{2}{*}{\textbf{Comment Text}} & \multicolumn{3}{c|}{\textbf{True Label}} & \multicolumn{3}{c|}{\textbf{Predicted Label}} \\ \cline{2-7}
& \textbf{Subject} & \textbf{Vaccinated} & \textbf{Regret} & \textbf{Subject} & \textbf{Vaccinated} & \textbf{Regret} \\
\hline
blah blah blah bu blah bu bu blah\dots guess who i am? i'm a cuflu shot recipient! gessssshhh when are going to recall all this crap and accept the fraud\dots & self & 1 & 0 & self & 1 & 1 \\
\hline
i bet he regrets getting the clot shot and the 30 boosters following. & other & 1 & 0 & other & 1 & 1 \\
\hline
crony liars and misinforms, million of people after the booster got tested positive to covid 19, does not protect from getting the virus, i am one of them , the main question is why we need the mandate & self & 1 & 0 & self & 1 & 1 \\
\hline
\end{tabular}
\end{table*}

\begin{table*}[htbp]
\centering
\caption{Examples of False Negatives (Regret)}
\label{tab:false_negative_examples}
\begin{tabular}{|p{0.4\textwidth}|ccc|ccc|}
\hline
\multirow{2}{*}{\textbf{Comment Text}} & \multicolumn{3}{c|}{\textbf{True Label}} & \multicolumn{3}{c}{\textbf{Predicted Label}} \\
& \textbf{Subject} & \textbf{Vaccinated} & \textbf{Regret} & \textbf{Subject} & \textbf{Vaccinated} & \textbf{Regret} \\
\hline
i give kudo's to this man .. i am in the same position forced to get the shot..got the shot still ended up getting covid..my job said no shot no paycheck & self & 1 & 1 & self & 1 & 0 \\
\hline
no one's died from the vaccine can you really be that ignorant and say that with a straight face marvin haggler there was a dr developed a rare blood disease and died he had a elite group of drs attempting to save his life but they could not and he said right before he died and this man is a dr who was a proponent for vaccines and he told his wife that it was no doubt the vaccine that killed him and there are plenty more to to go find & other & 1 & 1 & other & 1 & 0 \\
\hline
o.k. so, another booster, on top of the booster i got after my first booster six months after my initial 2 shot vaccination. what's next? wake me when the "bumblebee" variant shows up. c'mon man. & self & 1 & 1 & self & 1 & 0 \\
\hline
\end{tabular}
\end{table*}

Other notable error patterns involved the misclassification of Subject and Vaccinated status. The model sometimes struggles to differentiate between a general, non-specific group (e.g., "soooo many vaccinated") and a specific third-party group known to the commenter, incorrectly assigning an "other" label to a broad, impersonal observation. Furthermore, the model occasionally failed to infer vaccination status in complex or ambiguous edge cases. Table~\ref{tab:other_error_examples} contains examples of these misclassifications.

\begin{table*}[htbp]
\centering
\caption{Subject and Vaccinated misclassification examples}
\label{tab:other_error_examples}
\begin{tabular}{|p{0.4\textwidth}|ccc|ccc|}
\hline
\multirow{2}{*}{\textbf{Comment Text}} & \multicolumn{3}{c|}{\textbf{True Label}} & \multicolumn{3}{c}{\textbf{Predicted Label}} \\
& \textbf{Subject} & \textbf{Vaccinated} & \textbf{Regret} & \textbf{Subject} & \textbf{Vaccinated} & \textbf{Regret} \\
\hline
soooo many vaccinated realizing they've made a huge mistake & unspecified & 0 & 0 & other & 0 & 0 \\
\hline
stupid parents almost kill their child with a covid jab. parents get your head out of your asses! & other & 1 & 0 & other & 0 & 0 \\
\hline
how about the vaccine caused his med. problems & other & 1 & 0 & other & 0 & 0 \\
\hline
\end{tabular}
\end{table*}

\section{Crowd-sourced Annotation Study}
\label{app:compensation}

Our annotator pool consisted entirely of U.S. residents, with a near-equal gender split (106 male, 95 female) and an average age of approximately 42.5 years. Political affiliation was evenly distributed among Democrats, Republicans, and Independents (67 each). Most annotators were employed full-time (89), with others reporting part-time (31) or non-paid roles such as homemakers or retirees (23). The majority identified as White (141), followed by Black (28), Asian (17), and Mixed (12) ethnicities. The median time taken to complete a batch was around 19 minutes. A significant proportion were not students (128), and all participants were based in the United States.





\section{Computational Resources}
The models were finetuned on a H200 GPU for over 130 GPU Hours combined. Rest of the inferencing were done on 4 x A100 GPUs. 


\endgroup

\end{document}